\documentclass[12pt,draftclsnofoot,onecolumn]{IEEEtran}

\IEEEoverridecommandlockouts

\usepackage{graphicx}
\usepackage{subfigure}
\usepackage{amsmath}
\usepackage{amssymb}
\usepackage{setspace}
\usepackage[compress]{cite}
\usepackage{theorem}

\theorembodyfont{\rmfamily}
\newtheorem{theorem}{Theorem}

\newtheorem{remark}{Remark}

\makeatletter
\renewcommand{\citepunct}{,\penalty\@m\hskip.13emplus.1emminus.1em}
\renewcommand{\citedash}{\hbox{--}\penalty\@m}
\makeatother

\begin{document}
%\begin{CJK*}{GBK}{song}、
%\doublespacing
% paper title
\title{Impact of Channel Asymmetry on Performance of Channel Estimation and Precoding for Downlink Base Station Cooperative Transmission}

\author
{Xueying Hou, \IEEEmembership{Student Member, IEEE}, Chenyang Yang, \IEEEmembership{Senior Member, IEEE}, and  Buon Kiong (Vincent) Lau, \IEEEmembership{Senior Member, IEEE}

{\thanks{ X. Hou and C. Yang are with the School of Electronics and Information Engineering, Beihang University (BUAA), Beijing, 100191, China (e-mail: hxymr@ee.buaa.edu.cn; cyyang@buaa.edu.cn). B. K. Lau is with the Department of Electrical and Information Technology, Lund University, SE-221 00 Lund, Sweden (e-mail: bkl@eit.lth.se).}}}

\maketitle

\begin{abstract}
Base station (BS) cooperative transmission can improve the spectrum
efficiency of cellular systems, whereas using which the channels
will become \textit{asymmetry}. In this paper, we study the impact
of the \textit{asymmetry} on the performance of channel estimation
and precoding in downlink BS cooperative multiple-antenna multiple-carrier
systems. We first present three linear estimators which
jointly estimate the channel coefficients from users in different
cells with minimum mean square error, robust design and least square
criterion, and then study the impact of \textit{uplink channel
asymmetry} on their performance. It is shown that when the large
scale channel information is exploited for channel estimation, using
non-orthogonal training sequences among users in different cells
leads to minor performance loss. Next, we analyze the impact of
\textit{downlink channel asymmetry} on the performance of precoding
with channel estimation errors. Our analysis shows that although the
estimation errors of weak cross links are large, the resulting rate
loss is minor because their contributions are weighted by the
receive SNRs. The simulation results verify our analysis and show
that the rate loss per user is almost constant no matter where the user is
located, when the channel estimators exploiting the large scale
fading gains.
\end{abstract}
\section{Introduction} \label{S:Intro.}
Base station (BS) cooperative transmission, which is also known as
coordinated multi-point transmission (CoMP), is an effective way to
mitigate the inter-cell interference (ICI) arisen from universal
frequency reuse cellular systems. As a promising transmit strategy,
coherent cooperative transmission can enhance the downlink spectrum
efficiency by using multiuser (MU) multiple-input multiple-output
(MIMO) precoding \cite{Foschini-NetworkMIMO-06,Tolli08}, when both
data and channel state information (CSI) are gathered at a central
unit (CU) via backhaul links.

In non-cooperative systems, each BS only needs to estimate the CSI
of \textit{local channels}, \emph{i.e.}, the channels between the BS
and the mobile stations (MSs) that are in the same cell. If the
training sequences for the MSs in different cells are not
orthogonal, the channel estimation performance will severely degrade
due to the ICI
\cite{J.Jose-PilotContam-ISIT09,Kang-CellularPilot-Letter07,D.Katselis-TSunderCorrCH-08TSP,S.Lee-VarPilotDens-TWC08,M.R.Raghavendra-InfRej-TVT09}.
The impact of the ICI can be mitigated by designing training
sequences with low cross-correlation for the MSs in different cells
\cite{Kang-CellularPilot-Letter07}, or by developing channel
estimators exploiting the interference statistics
\cite{D.Katselis-TSunderCorrCH-08TSP}. In
\cite{S.Lee-VarPilotDens-TWC08}, the authors propose to use
non-uniform pilot density and a DFT-based channel estimator to first
separate and then subtract the interference signal from the
estimated channel impulse response (CIR). Assuming that the desired
channels and the interfered channels do not overlap and their
interference-free initial estimates can be obtained through
orthogonal training, the authors in
\cite{M.R.Raghavendra-InfRej-TVT09} propose to exploit the delay
subspace structure to improve the estimation performance of the
desired channels.

In coherent cooperative transmission systems, the CSI of
\textit{cross channels}, i.e., the channels between the BSs and the
MSs who are in different cells, needs to be estimated as well. Both
the local and cross channel coefficients can be jointly estimated
using the conventional estimators such as those in
\cite{LiYe-OptimalTS-TC02} when the training signals are orthogonal
both for the MSs within a cell and for the MSs among the coordinated
cells. However, large overhead is inevitable if orthogonal training
signals are used for all MSs in the cooperative cell cluster.
Moreover, this demands inter-cell signalling and protocol to
coordinate the training sequences \cite{CSISounding-ICC10}. Such a
burden will become more noticeable when the cooperative clusters are
formed in a dynamic way \cite{A.Papadogiannis-DynamiCluster-08ICC}.
In \cite{L.Thiele-VirtualPilots-VTC08}, the authors suggest to
spread the orthogonal sequences from slot to slot, which may lead to
outdated CSI at the transmitter under time-varying channels.
Considering the propagation delay differences in multicell channels,
a group of orthogonal training sequences that are robust to the
delay are designed in \cite{T.Kwon-PilotDelay-07VTC}, but the number
of sequences in the group is limited.

An inherent feature of the channels in CoMP systems is
\textit{asymmetry}. On one hand, the multi-cell downlink channels
are \textit{asymmetric}, which means that the average channel gains
from different BSs to one MS are different. On the other hand, the
multi-cell uplink channels are also \textit{asymmetric}, which means
that the average channel gains from MSs in different cells to one BS
differ. Such an asymmetric channel feature is fundamental in CoMP
systems, since the difference of the large scale fading gains cannot
 be compensated by an uplink or downlink power control mechanism. Specifically, if
the MSs in different cells compensate their large scale fading gain
differences towards one BS by power control, their receive signal
energy differences towards other BSs will increase. This is
analogous to the interference asynchrony feature, which cannot be
dealt with by time-advanced techniques \cite{Zhang-Molish-async08}.

In this paper, we study the impact of the channel \textit{asymmetry}
on the performance of joint channel estimators and on the
performance of downlink BS cooperative MIMO orthogonal frequency-division multiplexing (OFDM) systems with channel estimation errors.

Firstly, we introduce three joint estimators requiring different
channel statistics, which are the minimum mean square error (MMSE)
estimator, a robust estimator and the least square (LS) estimator.
We analyze the performance of these estimators when the
\textit{uplink channel asymmetry} is exploited. Our analysis shows
that if the training sequences are not orthogonal among cells, the
LS estimator will perform significantly worse than using orthogonal
sequences. On the other hand, the MMSE and robust estimators have
minor performance loss from those using orthogonal sequences, thanks
to the large attenuation of the cross channels.

Secondly, we analyze the impact of channel estimation errors on the
performance of CoMP\footnote{There are various transmission
strategies for CoMP transmission such as coherent and non-coherent transmission. For simplicity, we refer the coherent BS cooperative
transmission using MU MIMO precoding as CoMP transmission in the
following.} system using zero forcing beamforming (ZFBF) by deriving
the rate loss led by the channel estimation errors. At the first
glance, the cross channels that experience large path loss are hard
to estimate in practice since the transmission power at the MS side
is limited \cite{Tolli08, HY09}, which may degrade the downlink
transmission performance. Nonetheless, our analysis shows that when
the training sequences are orthogonal and the joint MMSE estimator
is applied, the contribution of channel estimation errors to the
rate loss is weighted by the receive SNR of the corresponding
channel link. As a result, even though the channel estimation errors
of cross channels are large, their impact on the rate loss is minor
owing to the fact that the receive SNRs of the cross links are
considerably lower than the local link. Interestingly, simulation
results demonstrate that the rate loss of MS is nearly invariant no
matter if the MS is located at the cell edge or cell center.

The rest of the paper is organized as follows. Section II introduces
the system and channel models. Section III and IV respectively
present three joint channel estimators and analyze their
performance. In Section V, we analyze the impact of channel
estimation errors on downlink CoMP transmission. Simulation results
are provided in Section VI to verify our analysis and to evaluate
the system performance. The paper is concluded in Section VII.

\textit{Notations:} Boldface upper and lower case letters
$\mathbf{X}$ and $\mathbf{x}$ represent matrices and vectors, and
standard lower case letters $x$ denote scalars. $\mathbf{X}^T$, $\mathbf{X}^H$ and $tr\{\mathbf{X}\}$ denote the transpose, Hermitian conjugate transpose and the trace of
$\mathbf{X}$. $\mathbf{X}(i,i)$, $\mathbf{X}(i,:)$ and $\mathbf{X}(:,i)$ represent the $(i,i)$th element, the $i$th row and the $i$th column of $\mathbf{X}$, respectively. $\|\mathbf{x}\|$ represents the
two-norm of $\mathbf{x}$, and $\mathrm{diag}\{\mathbf{x}\}$ is a
diagonal matrix with its elements.
$\mathbb{E}\{x\}$ is the expectation of a random
variable $x$. $\Re\{x\}$ and $|x|$ stand for the real part and the
norm of a complex scalar $x$. $\lfloor x \rfloor$ denotes
the largest integer no larger than a real number $x$ and $\lceil x \rceil$ represents the smallest integer no smaller than $x$.
Finally, $\mathbf{I}_N$ denotes
the identity matrix of size $N$, and $\mathbf{0}$ denotes the matrix
of zeros.

\section{System and Channel Models} \label{S:System_Model}
\subsection{BS Cooperative Transmission System and Channel Models}
Consider a centralized CoMP system, where $B$ BSs each equipped with
$N_t$ antennas cooperatively serve $M$ single-antenna MSs. We
consider time division duplexing (TDD) systems, where the CSI
required for MU MIMO precoding is obtained through uplink training
by exploiting the channel reciprocity. In the uplink training phase,
all MSs send training sequences and each BS estimates the CSI from
all MSs to it. Then the BSs forward the estimated CSI to the CU via
low latency backhaul links. The CU computes the precoding and then
sends back the precoding vectors to each BS for downlink
transmission.

We consider frequency selective channels. The channel is assumed to
be quasi-static, which means that the channel remains constant
during the uplink training and the downlink transmission.
The composite CIR from MS $m$ to antenna $a$ of BS $b$ can be expressed
as $\mathbf{g}^t_{m,b,a} = \alpha_{m,b}\mathbf{h}^t_{m,b,a}$, where
$\mathbf{g}^t_{m,b,a} = [g^t_{m,b,a}(0),\cdots,g^t_{m,b,a}(L-1)]^T
\in \mathbb{C}^{L \times 1}$, $\alpha_{m,b}$ is the large scale
fading coefficient including path loss and shadowing,
$\mathbf{h}^t_{m,b,a} = [h^t_{m,b,a}(0),\cdots,h^t_{m,b,a}(L-1)]^T
\in \mathbb{C}^{L \times 1}$ is the small scale fading channel
vector, $h^t_{m,b,a}(l)$ is the fading coefficient of the $l$th
resolvable path, which is a complex Gaussian random variable with
zero mean and variance $\sigma_{h_l}^2$, and $L$ is the number of
resolvable paths. We assume that
$\sum_{l=0}^{L-1}{\sigma_{h_l}^2}=1$.

\subsection{Uplink Training Phase}
Except that the uplink channels are asymmetric, the signal received
at one BS from MSs in different cells are asynchronous in CoMP
systems \cite{Zhang-Molish-async08}. Denote the propagation delay
from MS $m$ to BS $b$ as $\tau_{m,b}$. We assume that the cyclic
prefix in the OFDM symbol is long enough, such that the propagation
delays turn into phase shifts in the frequency domain channels.

Consider that all $M$ MSs in the cooperative cluster send training
sequences during the same uplink training duration. Denote the
frequency domain training sequence of the $m$th MS as $\mathbf{t}_m
= [t_m(0),\cdots,t_m(K-1)]^T  \in \mathbb{C}^{K \times 1}$, its
transmit power at each subcarrier as $p^u_m$, then the received
signal of the $k$th subcarrier at antenna $a$ of BS $b$ can be
expressed as
\begin{align} \label{E:ULRxSignal_PC}
r_{b,a}(k) =  \sum_{m=1}^M \sqrt{p^u_m} t_m(k)  g^f_{m,b,a}(k) +
n(k),
\end{align}
where $g^f_{m,b,a}(k) = \sum^{L-1}_{l=0}\alpha_{m,b}
h^t_{m,b,a}(l-\frac{\tau_{m,b}}{T_s}) \exp(-j \frac{2 \pi}{K} l k)$
denotes the composite channel frequency response (CFR) at the $k$th
subcarrier including both large scale fading and phase
shift led by propagation delay, $T_s$ is the sampling period,
$K$ is the subcarriers number of OFDM system, $h^f_{m,b,a}(k)=\sum^{L-1}_{l=0}
h^t_{m,b,a}(l) \exp(-j \frac{2 \pi}{K} l k)$ represents the small
scale CFR at the $k$th subcarrier, $n(k)$ is the additive white
Gaussian noise (AWGN) with zero mean and variance $\sigma_n^2$.

\begin{figure}
\center
\includegraphics[width=0.6\textwidth]{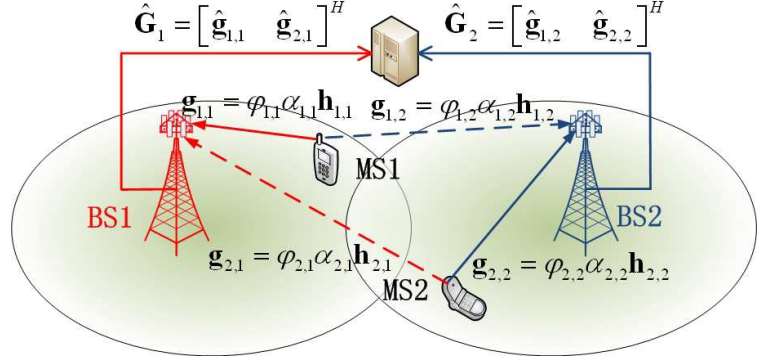}
\caption{\label{F:ULCoMP} Uplink training procedure of a two-cell
CoMP system, where the solid lines denote the local channels and the
dash lines represent the cross channels. The frequency domain
channels are shown and the index of subcarrier is omitted for
brevity.}
\end{figure}

With the received signal, BS $b$ estimates the composite CFR between
its multiple antennas and all MSs, $\mathbf{\hat{G}}^f_{b}(k) =
[\mathbf{\hat{g}}^f_{1,b}(k), \cdots, \mathbf{\hat{g}}^f_{M,b}(k)]^H
\in \mathbb{C}^{M \times N_t}$, where $\mathbf{\hat{g}}^f_{m,b}(k)=
[\hat{g}^f_{m,b,1}(k), \cdots, \hat{g}^f_{m,b,N_t}(k)]^T$ is the CFR
estimate between MS $m$ and the multiple antennas of BS $b$. Then
the BS forwards $\mathbf{\hat{G}}^f_{b}(k)$ to the CU, as
illustrated in Fig. \ref{F:ULCoMP} for a two-cell CoMP system.

\subsection{Downlink Transmission Phase} \label{S:DLModel}
\begin{figure}
\center
\includegraphics[width=0.6\textwidth]{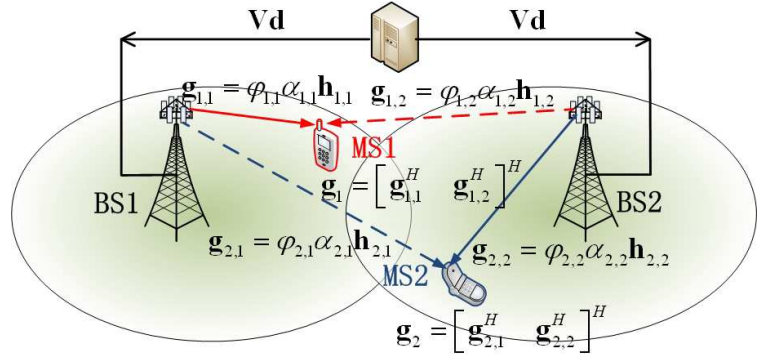}
\caption{\label{F:DLCoMP} Downlink transmission procedure of a two-cell CoMP system.}
\end{figure}

The CU collects the estimated composite CFR from each BS via a low
latency backhaul, and integrates the estimated global channel as
$\mathbf{\hat{G}}^f(k) = [\mathbf{\hat{G}}^f_{1}(k), \cdots,
\mathbf{\hat{G}}^f_{B}(k)] \in \mathbb{C}^{M \times BN_t}$. Then,
the CU computes the multi-cell precoding $\mathbf{V}(k)$ based on
$\mathbf{\hat{G}}^f(k)$, and sends the downlink data and the
corresponding precoder to each BS, as illustrated in Fig.
\ref{F:DLCoMP}.

The downlink composite channels from multiple BSs to MS $m$ can be
expressed as $\mathbf{g}^f_m(k) = [(\mathbf{g}^f_{m,1}(k))^H,
\cdots,(\mathbf{g}^f_{m,B}(k))^H]^H$, where $\mathbf{g}^f_{m,b}(k)
\in \mathbb{C}^{N_t \times 1}$ is the channel vector between all
antennas of BS $b$ and MS $m$. Let $d_m(k)$ be the data intended for
MS $m$ at the $k$th subcarrier. For simplicity and without loss of
generality, we assume that $\mathbb{E}\{d^*_m(k)d_m(k)\}=1$.
Denote $\mathbf{v}_m(k) \in \mathbb{C}^{BN_t \times 1}$ as the
precoding vector for MS $m$ under CoMP transmission, and $p_m^d(k)$
as the power allocated to MS $m$ at the $k$th subcarrier. Then the
receive signal at MS $m$ is
\begin{align} \label{E:DL_RxSignal}
    y_m(k) &= \sqrt{p_m^d(k)} (\mathbf{g}^f_m(k))^H \mathbf{v}_m(k) d_m(k) +   \sum_{j=1,j \neq m}^M  \sqrt{p_j^d(k)} (\mathbf{g}^f_m(k))^H \mathbf{v}_j(k) d_j(k)+ z_m(k),
\end{align}
where $z_m(k)$ is the AWGN with zero mean and variance $\sigma_z^2$ experienced at MS $m$.

\section{Uplink Channel Estimation for Downlink CoMP Transmission}
As shown in the downlink transmission model, the composite CFR is
required for precoding in CoMP OFDM systems, rather than the small
scale fading CFR.

The performance of channel estimation for the composite CFR depends
both on the channel features and on the information known \textit{a
priori}. In this paper, we assume that the propagation delays and
the number of resolvable paths can be estimated perfectly. In
practice, they can be estimated using various techniques such as those
shown in \cite{RG05,M.R.Raghavendra-InfRej-TVT09}. Since the number
of resolvable paths is usually much less than the number of
subcarriers in practical systems, the performance of channel
estimation can be significantly improved by exploiting the frequency
correlation of the channels \cite{LiYe-OptimalTS-TC02}. When the
propagation delays and the number of resolvable paths are known,
this can simply be implemented by first estimating the composite
CIR, and then obtaining the CFR by Fourier transformation. In the
following, we only address the CIR estimation.

To simplify our analysis, we assume that the transmit power for each
MS's training sequence is equal. The frequency domain receive signal
in (\ref{E:ULRxSignal_PC}) can be rewritten as a more compact form
\begin{align} \label{E:ULRxSignal_Vec}
\mathbf{r}_{b,a} &= \sqrt{p^u} \sum_{m=1}^M  \mathbf{T}_m \mathbf{g}^f_{m,b,a} + \mathbf{n}
%= \sqrt{p^u} \sum_{m=1}^M  \mathbf{T}_m \mathbf{\Phi}_{m,b} \alpha_{m,b} \mathbf{h}^f_{m,b,a} + \mathbf{n}
= \sqrt{p^u} \sum_{m=1}^M  \mathbf{T}_m \mathbf{\Phi}_{m,b} \alpha_{m,b} \mathbf{F} \mathbf{h}^t_{m,b,a} + \mathbf{n}\nonumber\\
&=\sqrt{p^u}\sum_{m=1}^M  \mathbf{X}_{m,b} \mathbf{g}^t_{m,b,a} +
\mathbf{n} = \sqrt{p^u}\mathbf{X} \mathbf{g}^t_{b,a} + \mathbf{n},
\end{align}
where $\mathbf{r}_{b,a} = [r_{b,a} (0),\cdots,r_{b,a}(K-1)]^T$,
$\mathbf{T}_m =\mathrm{diag}\{\mathbf{t}_m\}$, $\mathbf{g}^f_{m,b,a}
= [ g^f_{m,b,a}(0),\cdots,g^f_{m,b,a}(K-1)]^T$ denotes the composite
CFR vector from MS $m$ to the $a$th antenna of BS $b$, $
\mathbf{\Phi}_{m,b} = \mathrm{diag}\{[\psi_{m,b}(0), \cdots,
\psi_{m,b}(K-1)]\}$, $\psi_{m,b}(k) = \exp({-j \frac{2 \pi}{K}}
\frac{\tau_{m,b}}{T_s} k)$, $\mathbf{F} \in \mathbb{C}^{K \times L}$
is the first $L$ columns of a $K \times K$ Fourier transform matrix,
$\mathbf{X}_{m,b} = \mathbf{T}_m  \mathbf{\Phi}_{m,b} \mathbf{F} \in
\mathbb{C}^{K \times L}$ is the equivalent training matrix of MS $m$
by considering the known $\tau_{m,b}$ and $L$, $\mathbf{X} =
[\mathbf{X}_{1,b}, \cdots,\mathbf{X}_{M,b}] \in \mathbb{C}^{K \times
ML}$ is an equivalent training matrix of all MSs,
$\mathbf{g}^t_{b,a} =
[(\mathbf{g}^t_{1,b,a})^T,\cdots,(\mathbf{g}^t_{M,b,a})^T]^T \in
\mathbb{C}^{ML \times 1}$ is the composite CIR vector from all MSs
to antenna $a$ of BS $b$. $\mathbf{n}$ is the AWGN vector with zero
mean and covariance matrix $\sigma^2_n \mathbf{I}_{K}$.

Since each BS needs to estimate both local and cross channels for
CoMP transmission, it is natural to estimate the CIRs from all MSs
jointly, i.e., to estimate $\mathbf{g}^t_{b,a}$. Denote the
estimation errors of $\mathbf{g}^t_{b,a}$ as
$\mathbf{\tilde{g}}^t_{b,a} =\mathbf{g}^t_{b,a} -
\mathbf{\hat{g}}^t_{b,a}$, and its MSE as $\mathrm{MSE}_{b,a} =
\mathbb{E}\{\|\mathbf{\tilde{g}}^t_{b,a}\|^2\} =
\mathbb{E}\{\sum_{m=1}^M \|\mathbf{\tilde{g}}^t_{m,b,a}\|^2\}$. The
 MMSE estimator can be readily derived from
(\ref{E:ULRxSignal_Vec}) by minimizing the $\mathrm{MSE}_{b,a}$, which yields,
\begin{equation} \label{E:MMSE_Est}
\hat{\mathbf{g}}^{t^{\mathrm{MMSE}}}_{b,a} = \left(\mathbf{X}^H \mathbf{X}+
\frac{\sigma^2_n}{p^u} \mathbf{R}_{b,a}^{-1}\right)^{-1}\mathbf{X}^H
\mathbf{r}_{b,a},
\end{equation}
where $\mathbf{R}_{b,a} = \mathbb{E}\{\mathbf{g}_{b,a}
\mathbf{g}_{b,a}^H\}$ is the covariance matrix of the channels from
all $M$ MSs to the $a$th antenna of BS $b$. Assume that the small
scale fading channels among different BS-MS links are uncorrelated.
Then $\mathbf{R}_{b,a} =
\mathrm{diag}\{[{\alpha^2_{1,b}\mathbf{R}_{1,b,a}},\dots,{\alpha^2_{M,b}\mathbf{R}_{M,b,a}}]\}$,
$\mathbf{R}_{m,b,a}$ is the covariance matrix of the small scale
fading channel vector from MS $m$ to the $a$th antenna of BS $b$.

Although both $\mathbf{R}_{m,b,a}$ and $\alpha^2_{m,b}$ vary slowly
and can be estimated in practice \cite{MYi-EstPDP-09TVT},
$\alpha^2_{m,b}$ is a scalar parameter that can be estimated more
accurately and requires a much lower feedback rate. When
$\alpha^2_{m,b}$ is estimated perfectly but $\mathbf{R}_{m,b,a}$ is
unknown, by assuming uniform power delay profile (PDP) for small
scale fading channels similarly to the conventional robust channel
estimation algorithms \cite{LiYe-RobustCE-TC98}, we obtain a robust
channel estimator,
\begin{equation} \label{E:Robust_Est}
\hat{\mathbf{g}}^{t^{\mathrm{robust}}}_{b,a} = \left(\mathbf{X}^H \mathbf{X} +
\frac{\sigma^2_n}{p^u} \mathbf{D}_{b,a}^{-1}\right)^{-1}\mathbf{X}^H \mathbf{r}_{b,a},
\end{equation}
where $\mathbf{D}_{b,a} =
\mathrm{diag}\{[{{\alpha^2_{1,b}}\frac{1}{L}\mathbf{I}_{L}},\dots,{{\alpha^2_{M,b}}\frac{1}{L}\mathbf{I}_{L }}]\}$.

When we know nothing more than $\tau_{m,b}$ and $L$, we can apply
the LS estimator as
\begin{equation} \label{E:LS_Est}
\hat{\mathbf{g}}^{t^{{\mathrm{LS}}}}_{b,a} = \left(\mathbf{X}^H
\mathbf{X}\right)^{-1}\mathbf{X}^H \mathbf{r}_{b,a}.
\end{equation}

\section{Performance Analysis of The Channel Estimators} \label{S:PerfAnalysis}
In this section, we analyze the performance of the joint channel
estimators. We derive the MSE of the composite CIR estimates. Then
we discuss the impact on the performance of the estimators when the
training sequences are \emph{orthogonal} or \emph{non-orthogonal}.

When more than two MSs send training sequences in the uplink, it is
nontrivial to obtain an explicit expression of the MSE of the CIR
estimate. For mathematical tractability, we consider a simple but
fundamental scenario, where $B$ multiple-antenna BSs cooperatively
serve two single-antenna MSs, i.e., $M=2$, and the two MSs are
located in two cells.

\vspace{-0.5cm}
\subsection{MSE of Three Estimators}
We first derive the estimation error covariance matrix of the CIRs
from all MSs to the $a$th antenna of the $b$th BS, $\mathbf{R}_{
\mathbf{\tilde{g}}_{b,a}} = \mathbb{E}\{\mathbf{\tilde{g}}^t_{b,a}
(\mathbf{\tilde{g}}^t_{b,a})^H\}$, and then we can get the MSE of
the CIR estimation as $\mathrm{MSE}_{m,b,a} =
\sum_{i=(m-1)L+1}^{mL}\mathbf{R}_{\mathbf{\tilde{g}}^t_{b,a}}(i,i)$.
 From (\ref{E:MMSE_Est}), (\ref{E:Robust_Est}) and (\ref{E:LS_Est}), the covariance matrix for the estimators can be
obtained as follows by applying the Woodbury matrix identity
\cite{MatrCompu},
\begin{subequations} \label{E:Covariance}
\begin{eqnarray}
&& \mathbf{R}_{
\mathbf{\tilde{g}}^t_{b,a}}^{\mathrm{MMSE}} =
\left( \mathbf{R}_{b,a}^{-1} + \frac{p^u}{\sigma^2_n}\mathbf{B}\right)^{-1},\label{E:MMSE_Cov}\\
&& \mathbf{R}_{ \mathbf{\tilde{g}}^t_{b,a}}^{\mathrm{robust}} =
\mathbf{\Delta}^{\mathrm{robust}} + \mathbf{R}_{ \mathbf{\tilde{g}}^t_{b,a}}^{\mathrm{MMSE}},
\label{E:Robust_Cov}\\
&& \mathbf{R}_{\mathbf{\tilde{g}}^t_{b,a}}^{\mathrm{LS}} =
\frac{\sigma^2_n}{p^u}\mathbf{B}^{-1} \label{E:LS_Cov},
\end{eqnarray}
\end{subequations}
where $\mathbf{B} =\mathbf{X}^H \mathbf{X} =\left(
\begin{array}{ccc}
\mathbf{P}_{1,1} & \mathbf{Q}_{2,1}^H \\
\mathbf{Q}_{2,1} & \mathbf{P}_{2,2}
\end{array} \right),$ $\mathbf{P}_{m,m} = \mathbf{X}_{m,b}^H\mathbf{X}_{m,b} = \mathbf{F}^H \mathbf{\Phi}_{m,b}^H \mathbf{T}_m^H \mathbf{T}_m \mathbf{\Phi}_{m,b} \mathbf{F}$
is the equivalent auto-correlation matrix of the training sequence
for MS $m$, $m = 1,2$, $\mathbf{Q}_{2,1} =
\mathbf{X}_{2,b}^H\mathbf{X}_{1,b}= \mathbf{F}^H
\mathbf{\Phi}_{2,b}^H \mathbf{T}_2^H \mathbf{T}_1
\mathbf{\Phi}_{1,b} \mathbf{F}$ is the equivalent cross-correlation
matrix of the training sequences of MS $1$ and MS $2$, and
%\begin{figure*}[!t]
%\begin{equation} \label{E:Delta}
$\mathbf{\Delta}^{\mathrm{robust}} =
\frac{\sigma^2_n}{p^u}(\mathbf{B} + \frac{\sigma^2_n}{p^u}
\mathbf{D}_{b,a}^{-1})^{-1}\mathbf{B}(\mathbf{I}_{ML}-\mathbf{R}_{b,a}\mathbf{D}_{b,a}^{-1})(\mathbf{B}
+ \frac{\sigma^2_n}{p^u} \mathbf{D}_{b,a}^{-1})^{-1}+
\mathbf{R}_{b,a}\mathbf{B}[(\mathbf{B} + \frac{\sigma^2_n}{p^u}
\mathbf{R}_{b,a}^{-1})^{-1} - (\mathbf{B} + \frac{\sigma^2_n}{p^u}
\mathbf{D}_{b,a}^{-1})^{-1}]$.
%\end{equation}
%\hrulefill
%\end{figure*}

To further simplify our analysis and gain some insight into the
problem, we assume uniform PDP of the small scale fading channels.
Then the MMSE estimator degenerates to the robust estimator. From
(\ref{E:MMSE_Cov}), the MSE for MMSE estimator is derived as (see
Appendix \ref{A:proof-1} for details)
\begin{equation} \label{E:MSE_MMSE}
\mathrm{MSE}_{m,b,a}^{\mathrm{{MMSE}}} = \eta_{m,b}
\frac{\sigma^2_n}{p^u}\frac{1}{K} \sum\nolimits_{l=0}^{L-1}
f_{\mathrm{MMSE}}(\lambda_l),  ~m=1,2,
\end{equation}
where $\eta_{m,b} = 1/(1+\frac{\sigma_n^2}{
\alpha_{m,b}^2p^u}\frac{L}{K})$, $f_{\mathrm{MMSE}}(\lambda_l) =
\frac{1}{1-\beta \lambda_l^2}$, $\beta = 1/ \prod_{j=1}^2
(1+\frac{\sigma_n^2}{ \alpha_{j,b}^2p^u}\frac{L}{K})$ and
$\lambda_l^2$ is the $l$th eigenvalue of $\frac{\mathbf{Q}_{2,1}^H
\mathbf{Q}_{2,1}}{K^2}$, $0 \leq \lambda_l^2 < 1$.

From (\ref{E:LS_Cov}), the MSE for LS estimator can be derived as
(see Appendix \ref{A:proof-2} for details)
\begin{equation}\label{E:MSE_LS}
\mathrm{MSE}_{m,b,a}^{\mathrm{{LS}}} =
\frac{\sigma^2_n}{p^u}\frac{1}{K}
\sum\nolimits_{l=0}^{L-1}f_{\mathrm{LS}}(\lambda_l), ~m=1,2,
\end{equation}
where $f_{\mathrm{LS}}(\lambda_l) = \frac{1}{1-\lambda_l^2}$.

To minimize the MSE of the estimators, the matrix $\mathbf{B}$
should be a diagonal matrix \cite{I.Barhumi-OptimalPilot-TSP03}.
This requires that C1) $\mathbf{P}_{mm}$ is diagonal, which can be
satisfied when $\mathbf{T}_m^H \mathbf{T}_m = \mathbf{I}_K$, and C2)
$\mathbf{Q}_{2,1} = \mathbf{0}$, which demands the training
sequences of the MSs in two cells to be orthogonal.

The condition C1) holds when the training sequences have perfect
auto-correlation. When the virtual carriers (VC) in practical OFDM
systems are considered or the training sequences are not sent on all
subcarriers with equi-power and equi-spaced, C1) does not hold any
more.

To ensure C2), the training sequences for the MSs in different cells
should be orthogonal. This can be implemented in time or frequency
domain, but the resources occupied by training will increase
linearly with the number of MSs. Phase shift orthogonalization,
where two sequences are orthogonal when their relative phase shift
is larger than $\frac{2 \pi}{K}L$, is known as
an efficient way to generate training sequence with perfect
cross-correlation
\cite{LiYe-OptimalTS-TC02,I.Barhumi-OptimalPilot-TSP03}. At most
$\lfloor K/L \rfloor$ orthogonal sequences can be constructed from a sequence.

Considering the propagation delay in multi-cell scenarios, to
construct the phase shift orthogonal training sequences, the
relative phase shift of the two sequences for two MSs should exceed
$\frac{2 \pi}{K}\bar{L}$, where $\bar{L}=L + l_{delay}$ and
$l_{delay} = \lceil\frac{\tau_{m,b}-\tau_{i,b}}{T_s}\rceil$
represents the sampled propagation delay difference. Then, the
maximum number of orthogonal training sequences that can be
constructed is $\lfloor K/\bar{L}\rfloor$. When all MSs are located
in one cell, $l_{delay}$ is much smaller than $L$ in typical outdoor
channels\footnote{Take the urban macro channel in systems complying
Long Term Evolution (LTE) standard as an example, the multipath
delay is usually $4\sim5\mu$s. If we consider
 the cell radius to be $250$m, then the maximum delay difference is
$0.7\mu$s, which is negligible compared to the multipath delay.}. By
contrast, if the MSs are scattered in multiple cells, $l_{delay}$
will be comparable to the multipath delay. Consequently, few
orthogonal training sequences are available for a given sequence
length since $\bar{L}$ is large. This again leads to low spectrum
efficiency. Furthermore, the inter-cell orthogonality demands
inter-cell signalling and protocol to coordinate the training
resources\footnote{In practical cellular systems such as those
complying LTE standard, partial band may be used for uplink training to
increase the power spectrum density for cell edge MSs
\cite{LTE-TR36.211}. When there is no inter-cell coordination, the
training signals may overlap partially in the frequency domain,
which will increase the cross-correlation of the training signals
\cite{Huawei-SRS-58}.}, which will become a burden when the
coordinated clusters are formed dynamically.

In the following, we will analyze the performance loss led by the
non-orthogonal training. To highlight the impact of the
non-orthogonal training sequences for the MSs in different cells, we
assume that the condition C1) holds.

\vspace{-0.4cm}
\subsection{Impact of Non-Orthogonal Training}
For comparison, we first assume that the training sequences of MSs
in different cells are orthogonal, i.e.,
$\mathbf{Q}_{2,1}=\mathbf{0}$. Then the values of $\lambda_l$, $l=0,\cdots, L-1$, in
(\ref{E:MSE_MMSE}) and (\ref{E:MSE_LS}) are zeros, and we can see
the MSE for estimating the local and the cross channels. In this
case, both $f_{\mathrm{LS}}(\lambda_l)$  and
$f_{\mathrm{MMSE}}(\lambda_l)$ are equal to $1$, and the MSE of
 MMSE and LS estimators\footnote{Again we assume unform PDP in this subsection.}
 are
\begin{subequations}
\begin{eqnarray}
&&\mathrm{MSE}^{{\mathrm{MMSE}}}_{m,b,a} = \frac{1}{1+\frac{\sigma_n^2}{ \alpha_{m,b}^2p^u}\frac{L}{K}} \frac{\sigma^2_n}{p^u}\frac{L}{K} \label{E:RMSE_MMSE},\\
&&\mathrm{MSE}^{{\mathrm{LS}}}_{m,b,a} = \frac{\sigma^2_n}{p^u}\frac{L}{K}  \label{E:RMSE_LS}
.
\end{eqnarray}
\end{subequations}

For the LS estimator, the MSE of the composite CIR estimate,
$\mathbf{\hat{g}}^t_{m,b,a}$, depends on $\sigma^2_n$. We assume
that the noise variance at all BSs are the same. Then the MSE of
$\mathbf{\hat{g}}^t_{m,b,a}$ for $b=1,\cdots, B$ are identical, no
matter they are the local or the cross channels of MS $m$. For the
MMSE estimator, the MSE of $\mathbf{\hat{g}}^t_{m,b,a}$ depends on
$\alpha_{m,b}^2$ as well, which is the large scale fading energy of
$\mathbf{g}^t_{m,b,a}$. If MS $m$ is in the same cell as BS $c_m$,
then $\mathbf{g}^t_{m,c_m,a}$ is the local composite channel for MS
$m$ while $\mathbf{g}^t_{m,b,a}$ for $b \neq c_m$ are its cross
composite channels. Since the local channel energy
$\alpha_{m,c_m}^2$ is usually larger than the cross channel energy
$\alpha_{m,b}^2$, $b \neq c_m$, we can observe from
(\ref{E:RMSE_MMSE}) that the MSE of the weak cross channels is even
less than that of the strong local channels.

At the first glance, this conclusion is inconsistent with the
conventional understanding, where the MSE of the estimates of the
cross channels should be larger than that of local channels.
Nevertheless, this understanding is only applicable for estimating
the small scale fading channels whose average energy is 1. To see
this, we normalize the MSE of $\mathbf{\hat{g}}^t_{m,b,a}$ by
$\alpha_{m,b}^2$ to obtain a normalized MSE (NMSE) of
$\mathbf{\hat{g}}^t_{m,b,a}$, which is actually the MSE for
estimating the small scale fading channel
$\mathbf{\hat{h}}^t_{m,b,a}$. The NMSE for both MMSE and LS
estimators can be expressed as follows
\begin{subequations}
\begin{eqnarray}
&&\mathrm{NMSE}^{{\mathrm{MMSE}}}_{m,b,a} = \frac{1}{1+\frac{\sigma_n^2}{ \alpha_{m,b}^2p^u}\frac{L}{K}} \frac{\sigma^2_n}{\alpha_{m,b}^2p^u}\frac{L}{K} \label{E:NMSE_MMSE},  \label{E:NMSE_LS}\\
&&\mathrm{NMSE}^{{\mathrm{LS}}}_{m,b,a} = \frac{\sigma^2_n}{\alpha_{m,b}^2p^u}\frac{L}{K}.
\end{eqnarray}
\end{subequations}
It follows that the MSE of the estimates for the small scale fading
channels with low receive energy is larger than that with high
receive energy.

When the training sequences of MSs in different cells are not
orthogonal, then $ \lambda^2_l \neq 0$, and both
$f_{\mathrm{LS}}(\lambda_l)$ and $f_{\mathrm{MMSE}}(\lambda_l)$
exceed $1$.

From the expression of $f_{\mathrm{LS}}(\lambda_l)$, we can see that
if $\lambda^2_l$ is close to $1$ for any $l$, its value will be
extremely large and the estimation performance will be severely
degraded. This means that the LS estimator is quite sensitive to the
orthogonality of the training sequences.

In the expression of $f_{\mathrm{MMSE}}(\lambda_l)$, $\lambda^2_l$
is weighted by $\beta$, whose value is always less than $1$. If two
MSs are all in the same cell, $\alpha_{j,b}^2$ is generally large,
then $\beta$ is close to $1$ and $f_{\mathrm{MMSE}}(\lambda_l)
\approx f_{\mathrm{LS}}(\lambda_l)$. This implies if the MSs in the
same cell use non-orthogonal training sequences, the performance of
the MMSE estimator will degrade severely. On the other hand, if two
MSs are in different cells, since the large scale channel gains from
MSs to their non-serving BSs are low that leads to small $\beta$,
$f_{\mathrm{MMSE}}(\lambda_l)$ will not be too large even if
$\lambda_l^2$ is close to 1. This indicates that the MMSE estimator
is robust to the non-orthogonality of the training sequences in
different cells, thanks to the severe energy attenuation of the
channels from MSs to their non-serving BSs.

Note that this conclusion holds for both MSE and NMSE since they
only differ in a constant.

\subsection{Performance Gap between MMSE Estimator and Robust Estimator}
In wideband cellular systems, the PDP is in fact not uniform. Then
the robust estimator will be inferior to the MMSE estimator.
Nevertheless, we will show in the following analysis that the
performance gap between the two estimators is minor when the
training sequences are orthogonal. We will show through simulations
in Section \ref{S:Simulation} that the same conclusion can be drawn
when the training sequences are not orthogonal.

When the training sequences are orthogonal, $\mathbf{B} =
\mathbf{X}^H \mathbf{X}= K\mathbf{I}_{2L}$. Substituting
$\mathbf{B}$ into (\ref{E:MMSE_Cov}) and (\ref{E:Robust_Cov}), we
can derive the MSE difference of the MMSE estimator and robust
estimator as
\begin{align} \label{E:Delta_MMSE}
\Delta^{\mathrm{MMSE}}_{m,b,a} & = \sum_{i=(m-1)L+1}^{mL}
\left[\mathbf{R}_{ \mathbf{\tilde{g}}_{b,a}}^{\mathrm{robust}}(i,i) -
\mathbf{R}_{\mathbf{\tilde{g}}_{b,a}}^{\mathrm{MMSE}}(i,i)\right]
 = \alpha_{m,b}^2 \left(\sum_{l=0}^{L-1}\frac{\sigma_{h_l}^4}{\sigma_{h_l}^2+\mu} - \frac{1}{1+\mu L}\right),
\end{align}
where $\mu = \frac{\sigma_n^2}{\alpha^2_{m,b} p^u K}$,
$\sigma_{h_l}^2$ is the variance of the $l$th resolvable path of small
scale fading channel.

When $\alpha_{m,b}^2$ is large enough, $\mu$ approaches to $0$ and
$\Delta^{\mathrm{MMSE}}_{m,b,a}$ approaches to $0$ also. On the
other hand, when $\alpha_{m,b}^2$ decreases, $\mu$ will increase,
but $\Delta^{\mathrm{MMSE}}_{m,b,a}$ will still be fairly small.
That is to say, in the asymmetric channels, the robust estimator
performs closely to the MMSE estimator.

\vspace{-0.4cm}
\section{Impact of Channel Estimation Errors on Cooperative Transmission} \label{S:DL_Analysis}
In this section, we first analyze the average per MS rate loss led
by the channel estimation errors with CoMP transmission using ZFBF.
Then, we obtain a lower bound of the average achievable rate when
the MMSE estimator is applied.

\vspace{-0.4cm}
\subsection{CFR Estimation Errors} \label{S:DL_Analysis_CEE}
Since composite CFR is required for precoding in OFDM systems, we
need to transform the MSE of CIR estimate to the MSE of CFR estimate
at each subcarrier, $\sigma^2_{e_{m,b,a}}(k)$, $k=0,\cdots,K-1$. Denote the sum MSE of
CFR at all subcarriers as
$\mathbb{E}\{\|\mathbf{\tilde{g}}^f_{m,b,a}\|^2\} =
\mathbb{E}\{\|\mathbf{g}^f_{m,b,a} -
\mathbf{\hat{g}}^f_{m,b,a}\|^2\}$. It can be obtained from the MSE
of CIRs provided in previous sections as follows,
\begin{align}
\mathbb{E}\{\|\mathbf{\tilde{g}}^f_{m,b,a}\|^2\} =
\mathbb{E}\{\|\mathbf{F}_{m,b} \mathbf{g}^t_{m,b,a} -
\mathbf{F}_{m,b} \mathbf{\hat{g}}^t_{m,b,a}\|^2\} =
\mathbb{E}\{\|\mathbf{F}_{m,b} \mathbf{\tilde{g}}^t_{m,b,a}\|^2\}
\stackrel{(a)} =K\mathrm{MSE}_{m,b,a},
\end{align}
where (a) comes from the fact that $\mathbf{F}_{m,b}^H
\mathbf{F}_{m,b} = \mathbf{F}^H \mathbf{\Phi}_{m,b}^H
\mathbf{\Phi}_{m,b} \mathbf{F} =K \mathbf{I}_{L}$. When the training
sequences of all MSs are orthogonal and the resolvable multipaths
are uncorrelated, the MSE of the CFR at each subcarrier can be
obtained as \cite{LiYe-RobustCE-TC98}
\begin{equation}
\sigma^2_{e_{m,b,a}}(k) =
\mathbb{E}\{\|\mathbf{\tilde{g}}^f_{m,b,a}\|^2\}/{K} =
\mathrm{MSE}_{m,b,a}, ~ k=0,\cdots, K-1.
\end{equation}

From (\ref{E:MSE_MMSE}) and (\ref{E:MSE_LS}) we know that the MSE of
the CIR estimates between all antennas of BS $b$ and MS $m$ are the
same, which results in $\sigma^2_{e_{m,b,a}} = \sigma^2_{e_{m,b}}$,
$a = 1, \cdots, N_t$. The composite CFR between BS $b$ and MS $m$ at
the $k$th subcarrier can be modeled as $ \mathbf{g}^f_{m,b}(k) =
\mathbf{\hat{g}}^{f}_{m,b}(k) + \mathbf{\tilde{g}}^{f}_{m,b}(k)$,
where $\mathbf{g}^f_{m,b}(k) = [g^f_{m,b,1}(k), \cdots,
g^f_{m,b,N_t}(k)]^H$, $\mathbf{\hat{g}}^{f}_{m,b}(k)$ is the
estimation of $\mathbf{g}^{f}_{m,b}(k)$ and
$\mathbf{\tilde{g}}^{f}_{m,b}(k)$ is the estimation error vector
whose covariance is $\sigma^2_{e_{m,b}}(k) \mathbf{I}_{N_t}$. Since
the transmission procedures of all subcarriers are same, the index
of subcarrier is omitted in the following for brevity.

\vspace{-0.3cm}
\subsection{Impact of Channel Estimation Errors on CoMP Transmission}
When the global channel vectors are reconstructed at the CU from the
estimates provided by all coordinated BSs, a multicell ZFBF is
computed as follows
\begin{equation} \label{E:precoder}
\mathbf{V} = (\mathbf{\hat{G}}^f)^H \left[ \mathbf{\hat{G}}^f (\mathbf{\hat{G}}^f)^H\right]^{-1}.
\end{equation}
Then the beamforming vector of all cooperative BSs for MS $m$ is
obtained by normalizing the $m$th column of $\mathbf{V}$ as
$\mathbf{v}_m = \mathbf{V}(:,m)/\|\mathbf{V}(:,m)\|$ and
$\mathbf{v}_m = [(\mathbf{v}_{1,m})^H, \cdots,
(\mathbf{v}_{B,m})^H]^H$, where $\mathbf{v}_{b,m} \in
\mathbb{C}^{N_t \times 1}$ is the precoder vector of BS $b$ for MS
$m$.

In order to derive a closed-form expression of the per MS rate loss
led by the channel estimation errors, we assume that the number of
MSs cooperatively served by $B$ BSs is $BN_t$, which indicates full
multiplexing CoMP-MU transmission as in \cite{Caire09}. We further
assume that the power allocated to all MSs are identical, which is
denoted as $p^d$.

The average rate of MS $m$ achieved by CSI estimate-based ZFBF is
obtained from \eqref{E:DL_RxSignal} as
\begin{equation}
R_m = \mathbb{E} \left\{ \log_2(1 + \mathrm{SINR}_m)\right\} = \mathbb{E}\left\{\mathrm{log}_2\left(1 +  \frac{p^d |\mathbf{g}_{m}^H \mathbf{v}_{m}|^2}{\sigma_z^2 +
p^d\sum_{j=1,j \neq m}^M { |\mathbf{g}_{m}^H \mathbf{v}_{j}}|^2}\right)\right\}.
\end{equation}

The average rate of MS $m$ achieved by perfect CSI-based ZFBF is
given by
\begin{equation}
R^{\mathrm{Ideal}}_m = \mathbb{E}\left\{\mathrm{log}_2\left(1 + \frac{p^d}{\sigma_z^2} |\mathbf{g}_{m}^H \mathbf{v}_{m}^{\mathrm{Ideal}}|^2\right)\right\},
\end{equation}
where $\mathbf{v}_{m}^{\mathrm{Ideal}} =
[(\mathbf{v}_{1,m}^{\mathrm{Ideal}})^H,
\cdots,(\mathbf{v}_{B,m}^{\mathrm{Ideal}})^H]^H$ is the perfect
CSI-based ZFBF vector of all BSs for MS $m$, which is chosen to
be orthogonal to $\mathbf{g}_{j}$ for $j=1,\cdots,M$ and $j \neq m$.

\begin{theorem} \label{C:RateGap}
\label{theorem:cmcash}

The rate loss of MS $m$ of the CoMP transmission using ZFBF led by
the channel estimation errors can be upper bounded by
\begin{equation} \label{E:RateGap}
\Delta R_m = R^{\mathrm{Ideal}}_m - R_m < \mathrm{log}_2\left(1 + \frac{p^d}{\sigma_z^2}\mathbb{E}\{I_m\}\right) = \Delta R_m^{\mathrm{UB}},
\end{equation}
where $I_m = \sum_{j=1, j \neq m}^M |(\mathbf{\tilde{g}}_m^f)^H
\mathbf{v}_j|^2$ is the average interference power experienced by MS
$m$, $\mathbf{\tilde{g}}_m^f = [(\mathbf{\tilde{g}}^{f}_{m,1})^H,
\cdots,(\mathbf{\tilde{g}}^{f}_{m,B})^H]^H$ is the estimation error
vector of the global channel vector $\mathbf{g}^{f}_{m}$.
\end{theorem}

The derivation is similar to that in \cite{Caire09}. Due to the lack
of space, we omit the proof of the Theorem. From \eqref{E:RateGap},
the achievable rate of MS $m$ when estimated CSI are used for CoMP
transmission can be lower bounded by
\begin{equation} \label{E:RateLB}
R_m > R^{\mathrm{Ideal}}_m - \Delta R_m^{\mathrm{UB}}.
\end{equation}

To gain further insight into the rate loss, we assume that the
channel estimation errors $\mathbf{\tilde{g}}_m^f$ are independent
of the precoder vectors $\mathbf{v}_j$ for $j=1,\cdots,M$ and $j
\neq m$. This assumption is satisfied when the MMSE estimator is
applied. When MMSE estimator is used, since the channel
estimation errors are independent of the channel estimates, and the
precoders are functions of the channel estimates, the channel
estimation errors and the precoders are mutually independent. In
Section \ref{S:Simulation}, we will examine the impact of estimation
errors led by the LS and robust estimators on CoMP transmission
through simulations.

As stated in Section \ref{S:DL_Analysis_CEE}, the MSEs of the CFRs
between MS $m$ and all antennas of BS $b$ are identical, then the
rate loss upper bound of MS $m$ can be further derived as follows by
taking expectation over the channel estimation errors (see
Appendix \ref{A:proof-4} for details)
\begin{align}\label{E:I_C1}
\Delta R_m^{\mathrm{UB}}
= \mathrm{log}_2[1 +
\sum_{j=1,j \neq m}^M  \sum_{b=1}^B \underbrace{\frac{\alpha^2_{m,b}
p^d}{\sigma_z^2}}_{\mathrm{SNR}_{m,b}^d}\underbrace{\frac{\sigma^2_{e_{m,b}}}
{\alpha^2_{m,b}}}_{\mathrm{NMSE}_{m,b}} \|\mathbf{v}_{b,j}\|^2].
\end{align}

\begin{remark}
To connect with the conventional understanding of the impact of the
channel estimation errors, we show the contribution of channel
estimation errors of the small scale fading channels,
$\mathrm{NMSE}_{m,b}$, to the rate loss in (\ref{E:I_C1}), which is
weighted by the downlink receive SNR of the link
$\mathrm{SNR}_{m,b}^d$. For the local channels of MS $m$, i.e.,
$b=c_m$, $\mathrm{NMSE}_{m,c_m}$ is small and its contribution to
the rate loss will be minor. For the cross channels of MS $m$, the
estimation errors $\mathrm{NMSE}_{m,b}$ for $b \neq c_m$ will be
large. However, because the receive SNRs of the cross links are
considerably low, the impact of the channel estimation errors of
cross channels will be significantly alleviated. This is true
especially for cell center MSs.
\end{remark}

Substituting the $\mathrm{NMSE}_{m,b}$ of the MMSE estimator under
orthogonal training shown in (\ref{E:NMSE_MMSE}) into
(\ref{E:I_C1}), we obtain the rate loss upper bound as follows
\begin{align}\label{E:I_MMSE_Orthog}
\Delta R_m^{\mathrm{UB}}
&= \mathrm{log}_2 \left[1 + \frac{p^d}{\sigma_z^2}\frac{\sigma^2_n}{p^u}\frac{L}{K}\sum_{j=1,j \neq m}^M  \sum_{b=1}^B\frac{\|\mathbf{v}_{b,j}\|^2}{1+\frac{\sigma_n^2}{ \alpha_{m,b}^2p^u}\frac{L}{K}} \right]
%&\stackrel{(a)}< \mathrm{log}_2\left[1 +  \frac{p^d}{\sigma_z^2}\frac{\sigma^2_n}{p^u}\frac{L}{K} \sum_{j=1,j \neq m}^M\sum_{b=1}^B\|\mathbf{v}_{b,j}\|^2 \right] \nonumber\\
\stackrel{(a)}< \mathrm{log}_2\left[1 + (M-1)
\frac{p^d}{\sigma_z^2}\frac{\sigma^2_n}{p^u}\frac{L}{K} \right],
\end{align}
where (a) is obtained because $\frac{1}{1+\frac{\sigma_n^2}{ \alpha_{m,b}^2p^u}\frac{L}{K}} < 1$,
and $\sum_{b=1}^B \|\mathbf{v}_{b,j}\|^2 = 1$ as described in \eqref{E:precoder}. $p^d$ and $p^u$ are
respectively the power transmitted to each MS and that transmitted
by each MS, $\sigma_z^2$ and $\sigma^2_n$ are the noise
variances at the BS and MS. When $p^d$ and $p^u$ are fixed, the
upper bound of the rate loss will not depend on the large scale
fading gains of both local and cross channels.

According to \eqref{E:RateLB} and (\ref{E:I_MMSE_Orthog}), we can
obtain the lower bound of the average rate achieved by MS $m$ under
MMSE estimator and orthogonal training as
\begin{equation} \label{E:LowerBound}
R_m > R^{\mathrm{Ideal}}_m - \mathrm{log}_2\left[1 + (M-1) \frac{p^d}{\sigma_z^2}\frac{\sigma^2_n}{p^u}\frac{L}{K} \right] = R_m^{\mathrm{LB}}.
\end{equation}

\section{Simulation Results} \label{S:Simulation}
In this section, we compare the performance of different channel
estimators and evaluate their impact on the performance of downlink
CoMP system.

A cooperative cluster of two cells is considered ($B = 2$). The cell
radius $r=250$ m. Each BS has four omnidirectional antennas ($N_t =
4$), serving two single-antenna MSs. The downlink transmit power for
each MS $p^d$ is $5$ dB larger than the uplink transmit power of
each MS $p^u$. The maximum delay spread of the channel $\tau = 4
\mu$s, and the channel is implemented as a tapped-delay line with
Rayleigh fading coefficients and an exponential DPD with the
attenuation factor being $1.4$. $K=128$. The system bandwidth is $B
= 5$ MHz, the sampling period is $T_s = 1/B= 0.2 \mu$s, thus
$L=\tau/T_s = 20$. Due to the lack of the space, we only present the
performance of an OFDM system without VC and the training sequences
are sent on full band. Extensive results show that the same
conclusions can be drawn to the OFDM system with VC and the training
sequences of MSs are sent on partial band.

The training sequences are constructed from Constant Amplitude Zero
Autocorrelation Code (CAZAC) \cite{ZCseq} as $t(k) = e^{-j \frac{\pi
c n_k(n_k+1)}{N_{\mathrm{ZC}}}}, ~k= 0,\cdots, K-1,$
\cite{LTE-TR36.211}, where $N_{\mathrm{ZC}}=127$, $n_k =
\mathrm{mod}(k,N_{\mathrm{ZC}})$. The training sequences for MSs in
the same cell are orthogonal by cyclic shifting. The training
sequences of MSs in different cells can be orthogonal or
non-orthogonal.  For orthogonal training, the training sequences for
the four MSs in the two cells are constructed from the cyclic shift
of the CAZAC with the same value of $c$. For non-orthogonal
training, the training sequences of MSs in two cells use different
values of $c$. The values of $c$ for two cells are set to $c_1 = 1$
and $c_2 = 7$, considering that their cross correlation is moderate.

\subsection{NMSEs of Different Estimators}
\begin{figure}
\center
\includegraphics[width=0.5\textwidth]{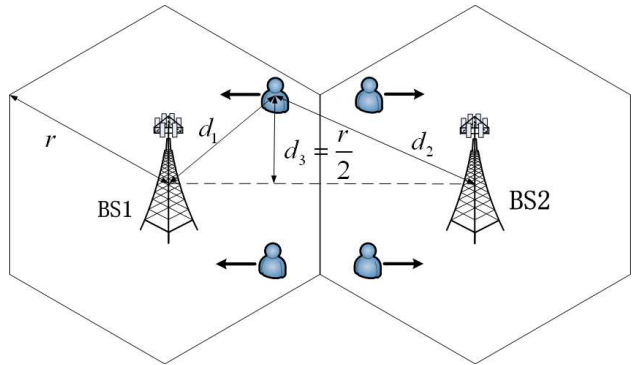}
\caption{\label{F:MSposition} MSs' positions in the simulated CoMP
system. The locations of two MSs in the same cell are symmetrical to
the line connecting the two BSs, and the MSs in different cells are
symmetric to the cell edge. $d_3$ is fixed to be $\frac{r}{2}$. All
MSs move from the cell edge to the cell center simultaneously, then
their local receive SNRs all increase. Given the value of $d_1$, we
can get the value of $d_2$ and vise versa. Assume that the downlink
receive SNR of the cell edge MS, $\mathrm{SNR}_{edge}$, is $10$dB.
Consider the path loss factor $\epsilon$ as $3.76$, then the
receive SNR of a MS from a BS with a distance $d$ can be computed
as $\mathrm{SNR}(d) = \mathrm{SNR}_{edge} +
\epsilon10\log_{10}(\frac{r}{d})$. Similarly, when $\mathrm{SNR}(d)$
is given, we can get $d$. }
\end{figure}

To show the impact of MSs' positions on the estimation errors for
small scale fading channels under orthogonal and non-orthogonal
training for multiple MSs, we let the four MSs in the two cells be
symmetrically located, as shown in Fig. \ref{F:MSposition}. Then the
channel estimation performance of all MSs are the same. We take the
performance of one MS as an example to analyze.

\begin{figure}
\center
\includegraphics[width=0.7\textwidth]{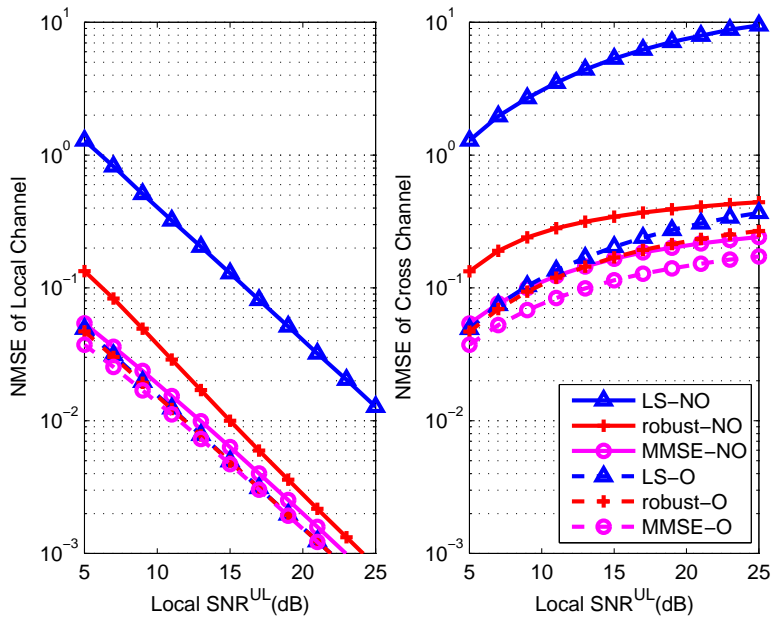}
\caption{\label{F:NMSE} NMSEs of different estimators for both local
and cross channels versus the receive SNR of the local channel. The
NMSEs are obtained by averaging over $1000$ realizations of small
scale fading channels. The X-axis is defined as
$\mathrm{Local~SNR^{UL}} = \frac{\alpha^2_{\mathrm{local}} p^u}{\sigma_n^2}$,
where $\alpha^2_{\mathrm{local}}$ is the large scale fading energy of the
local channel for the MS. When the MSs are at the cell edge,
$\mathrm{Local~SNR^{UL}} = 5$ dB. The Y-axis is the NMSE of channel
estimators, which reflects the estimation performance of the small
scale fading channels. When $\mathrm{Local~SNR^{UL}}$ increases,
the large scale fading gains of the cross channels decrease, which
leads to large NMSE of the cross channels. For the local channels,
the NMSE of the three estimators are overlapped under orthogonal
training (shown as "-O" in the legend). For the cross channels, the
NMSE of the MMSE estimator under non-orthogonal training  (shown as
"-NO" in the legend) is overlapped with that of the robust estimator
under orthogonal training.}
\end{figure}

In Fig. \ref{F:NMSE}, the NMSEs versus local uplink receive SNR of
three estimators for both local and cross channels are shown. When
the training sequences are not orthogonal, the performance of the LS
estimator degrades severely. By contrast, the performance gap of the
MMSE estimator under orthogonal and non-orthogonal training is
minor. Comparing the NMSE of the robust and MMSE estimators, we
observe that the performance loss of the robust estimator from the
MMSE estimator is small. These results agree well with our previous
analysis. Again, we should note that the impact of the
non-orthogonal training sequences on the performance of estimating
the small scale fading channels is the same as that of estimating
the composite channels, since the MSE and NMSE only differ in a
constant.

\subsection{Downlink Average Rate with Different Channel Estimators}
\subsubsection{Positions of MSs are Fixed}
Now we verify the analysis in Section V, where the positions of the
MSs in two cells are the same with before. We simulate the downlink
average throughput of each MS, which is averaged over $1000$
realizations of small scale fading channels.

\begin{figure}
\center
\includegraphics[width=0.6\textwidth]{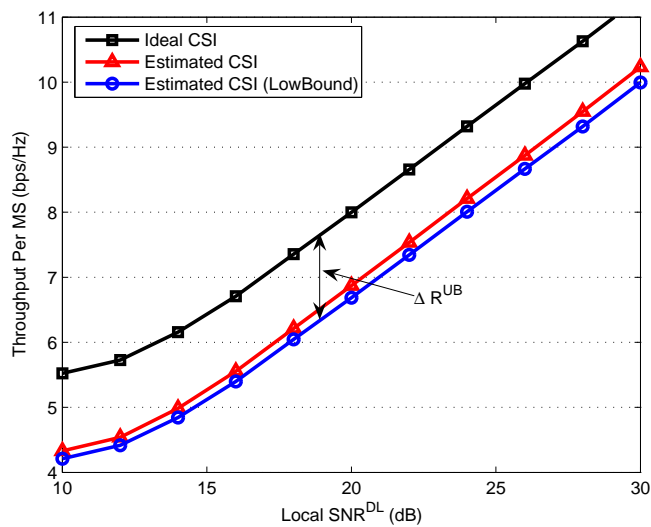}
\caption{\label{F:Thp_LB} Achievable rate and its lower bound of a
MS when the MMSE estimator and orthogonal training are considered.
The per MS rate of CoMP transmission with perfect CSI is also
provided for reference, which is shown as "ideal CSI" in the
legend.}
\end{figure}

We first evaluate the tightness of the rate lower bound derived in
Section \ref{S:DL_Analysis}. The performance with MMSE estimator
under orthogonal training is taken as an example. The per MS rate
under the pefect ZFBF and the CSI estimate-based ZFBF are shown in
Fig. \ref{F:Thp_LB}, together with the lower bound of the achievable
rate derived in (\ref{E:LowerBound}). It shows that the derived
lower bound is close to the rate obtained by simulation.

\begin{figure}
\center
\includegraphics[width=0.6\textwidth]{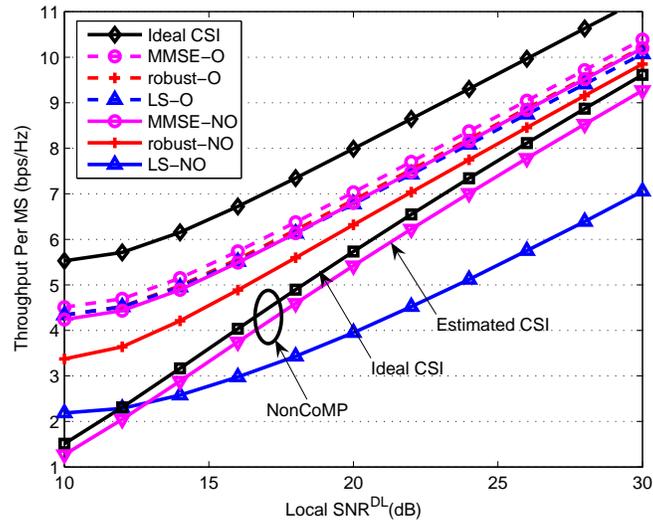}
\caption{\label{F:ThpSNR} Achievable rate of a MS when different
estimators are used with both orthogonal and non-orthogonal
training. For the Non-CoMP transmission with estimated CSI, the CSI
for downlink precoding is estimated by the conventional single user
MMSE estimator \cite{LiYe-OptimalTS-TC02}. The performance when the
robust estimator and the LS estimator under orthogonal training are
applied overlap with that when MMSE estimator under non-orthogonal
training are used. The meaning of the legends is the same as
previous figures.}
\end{figure}

We then compare the impact of different estimators on the
performance of CoMP transmission under both orthogonal and
non-orthogonal training, which is shown in Fig. \ref{F:ThpSNR},
where the throughputs under Non-CoMP transmission are also present
as a reference. When the training sequences are orthogonal, the per
MS throughputs under different estimators are almost the same. The
performance gap from the perfect CSI-based ZFBF does not change no
matter the MS is located at the cell edge or the cell center. When
the training sequences of MSs in different cells are not orthogonal,
the performance degradation when using both the MMSE and the robust
estimators is minor. On contrary, the performance when using the LS
estimator severely degrades, which is even worse than the Non-CoMP
transmission. It indicates that the estimator can perform fairly
well only when the large scale fading information is employed.

\subsubsection{Positions of MSs are Randomized}
\begin{figure}
\center
\includegraphics[width=0.7\textwidth]{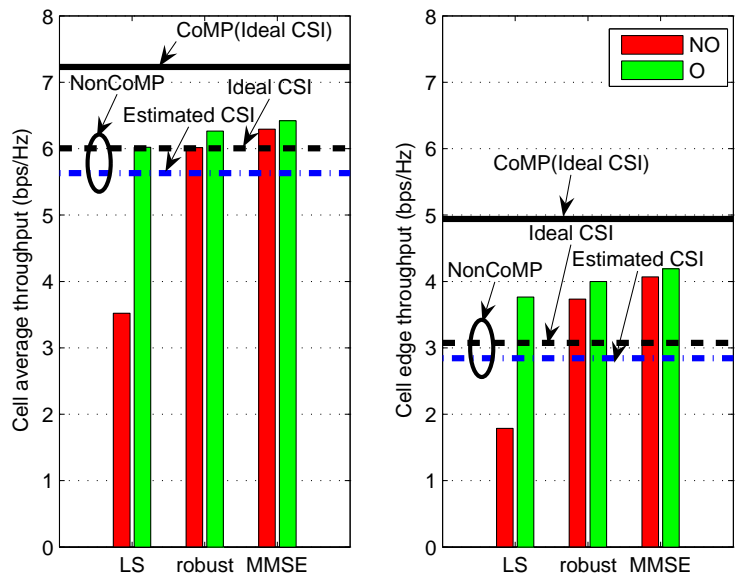}
\caption{\label{F:ThpBar} Cell average and cell edge MS throughput
when different channel estimators are used. The cell edge MS
throughput is defined as the $5\%$ point of the cumulative
distribution function of the MS throughput. As a baseline, the
results of Non-CoMP transmission are also provided. The meaning of the legends is
the same as previous figures.}
\end{figure}

Finally we simulate the case in which the locations of two MSs are
randomly distributed in each cell. We use the average throughput per MS and
the cell edge MS throughput as the performance metric. The results
are shown in Fig. \ref{F:ThpBar}, which is obtained from $1000$ random
drops. We can see that the same conclusion can be drawn as the case
where the positions of MSs are fixed.

\section{Conclusion}
In this paper, we have studied the impact of \emph{uplink channel
asymmetry} on the performance of channel estimation and the impact
of \emph{downlink channel asymmetry} on the performance of BS
cooperative transmission with channel estimation errors. We have
analyzed three joint channel estimators, the MMSE estimator, a
robust estimator and the LS estimator. Our analysis showed that if
the training sequences of MSs in different cells are not orthogonal,
the performance of estimating cross channels with the LS estimator
severely degrades. By contrast, the MMSE estimator is robust to the
non-orthogonal training, and due to the \emph{uplink channel
asymmetry} the robust estimator has minor performance loss. By
analyzing the impact of channel estimation errors on the cooperative
ZFBF transmission, we showed that due to the \emph{downlink channel
asymmetry}, the contribution of the channel estimation errors to the
rate loss is weighted by the receive SNRs of the corresponding
links. As a result, despite that the estimation errors of the small
scale fading channels of the cross links are large, their impact on
the rate loss is minor. When the joint channel estimators exploit
the large scale fading gains, CoMP transmission performs fairly
well, even without inter-cell orthogonal training. This improves the
spectral efficiency and simplify the inter-cell signalling required
to coordinate the training resources.

\appendices
\section{Derivation of the MSE of MMSE estimator}
\label{A:proof-1} Assume uniform PDP for small scale fading
channels, i.e., $\mathbf{R}_{b,a} =
diag\{[\alpha^2_{1,b}\frac{1}{L}\mathbf{I}_{L
},\alpha^2_{2,b}\frac{1}{L}\mathbf{I}_{L}]\}$. Substituting the
expression of $\mathbf{B}$ to (\ref{E:MMSE_Cov}) and applying the
formula of block matrix inversion, the covariance matrix of the
estimation errors becomes
\begin{equation} \label{E:Apdx_Cov_MMSE}
\mathbf{R}_{ \mathbf{\tilde{g}}_{b,a}}^{\mathrm{MMSE}}
=\frac{\sigma_n^2}{p^uK} \left(
\begin{array}{ccc}
\mathbf{N} &
- \eta_{b,2}\mathbf{N} \frac{\mathbf{Q}_{21}^H}{K}  \\
-  \eta_{b,2} \frac{\mathbf{Q}_{21}}{K} \mathbf{N} &
(\eta_{b,2})^{2}\frac{\mathbf{Q}_{21}}{K}\mathbf{N} \frac{\mathbf{Q}_{21}^H}{K} +\eta_{b,2}\mathbf{I}_L
\end{array} \right),
\end{equation}
where $\mathbf{N} = \left(\frac{1}{\eta_{b,1}}\mathbf{I}_L - \eta_{b,2}
\frac{\mathbf{Q}_{21}^H \mathbf{Q}_{21}}{K^2}\right)^{-1}$, $\eta_{b,m} =
\frac{\alpha^2_{b,m}}{\alpha^2_{b,m} + \frac{L\sigma^2_n}{K}}$.

The MSE of the MMSE estimator for CIR from the MS $1$ to antenna $a$
of BS $b$ is
\begin{align}\label{E:Apdx_MSE1_MMSE}
\mathrm{MSE}_{1,b,a}^{\mathrm{MMSE}}  =
\frac{\sigma_n^2}{p^uK}tr\{\mathbf{N}\}
&= \eta_{b,1}\frac{\sigma_n^2}{p^uK}tr \left\{\left(\mathbf{I}_L - \eta_{b,1}\eta_{b,2} \frac{\mathbf{Q}_{21}^H \mathbf{Q}_{21}}{K^2}\right)^{-1}\right\} \nonumber\\
%&= \eta_{b,1}\frac{\sigma_n^2}{p^uK}tr\{\mathbf{U}( \mathbf{I}_L- \eta_{b,1}\eta_{b,2} \mathbf{\Lambda} )^{-1}\mathbf{U}^H\} \nonumber\\
&\stackrel{(a)}= \eta_{b,1} \frac{\sigma_n^2}{p^uK} \sum_{l=0}^{L-1}{\frac{1}{1- \lambda_l^2 \prod_{j=1}^2 \eta_{b,j}}},
\end{align}
where (a) follows from applying the eigenvalue decomposition as
$\frac{\mathbf{Q}_{21}^H \mathbf{Q}_{21}}{K^2} = \mathbf{U}
\mathbf{\Lambda} \mathbf{U}^H$, $\mathbf{U}$ is an unitary matrix
and $\mathbf{\Lambda} = \mathrm{diag}\{[\lambda_0^2, \cdots,
\lambda_{L-1}^2]\}$.

The MSE of MMSE estimator for CIR from the MS $2$ can be derived
as
\begin{align}\label{E:Apdx_MSE2_MMSE}
\mathrm{MSE}_{2,b,a}^{\mathrm{MMSE}} = & \frac{\sigma_n^2}{p^uK}tr\left\{\eta_{b,2}^{2}\frac{\mathbf{Q}_{21}}{K}\mathbf{N} \frac{\mathbf{Q}_{21}^H}{K} +\eta_{b,2}\mathbf{I}_L\right\} \nonumber \\
=& \frac{\sigma_n^2}{p^uK}\eta_{b,2}^2tr\left\{\left(\frac{1}{\eta_{b,1}}\mathbf{I}_L - \eta_{b,2} \frac{\mathbf{Q}_{21}^H\mathbf{Q}_{21}}{K^2}\right)^{-1} \frac{\mathbf{Q}_{21}^H}{K}\frac{\mathbf{Q}_{21}}{K}\right\} +\frac{\sigma_n^2}{K}\eta_{b,2}tr\left\{\mathbf{I}_L\right\} \nonumber \\
%=& \frac{\sigma_n^2}{p^uK}(\eta_{b,2})^2\eta_{b,1}tr\{\mathbf{U}( \mathbf{I}_L - \eta_{b,1}\eta_{b,2} \mathbf{\Lambda} )^{-1}\mathbf{U}^H \mathbf{U}\mathbf{\Lambda}\mathbf{U}^H\}
%+\frac{\sigma_n^2}{K}\eta_{b,2}L \nonumber \\
=& \frac{\sigma_n^2}{p^uK}\eta_{b,2}^2\eta_{b,1}\sum_{l=0}^{L-1}{\frac{\lambda_l^2}{1- \lambda_l^2 \eta_{b,1}\eta_{b,2}}} +\frac{\sigma_n^2}{K}\eta_{b,2}L = \eta_{b,2} \frac{\sigma_n^2}{p^uK} \sum_{l=0}^{L-1}{\frac{1}{1-\lambda_l^2 \prod_{j=1}^2 \eta_{b,j} }}.
\end{align}

Comparing (\ref{E:Apdx_MSE1_MMSE}) and (\ref{E:Apdx_MSE2_MMSE}), we
can see that the only difference is the factor $\eta_{b,m}$. The
channel that experiences large attenuation exhibits small estimate
errors.

 \vspace{-0.4cm}
\section{Derivation of the MSE of LS estimator} \label{A:proof-2}
Substituting the expression of $\mathbf{B}$ into (\ref{E:LS_Cov})
and using the formula of $2\times2$ block matrix inversion, we can
derive the covariance matrix of the estimator errors as
\begin{equation} \label{E:Apdx_Cov_LS}
\mathbf{R}_{ \mathbf{\tilde{g}}_{b,a}}^{\mathrm{LS}}
=\frac{\sigma_n^2}{p^uK} \left(
\begin{array}{ccc}
\mathbf{M} &
-\mathbf{M} \frac{\mathbf{Q}_{21}^H}{K}  \\
- \frac{\mathbf{Q}_{21}}{K} \mathbf{M} &
\frac{\mathbf{Q}_{21}}{K}\mathbf{M} \frac{\mathbf{Q}_{21}^H}{K} + \mathbf{I}_L
\end{array} \right),
\end{equation}
where $\mathbf{M} = \left(\mathbf{I}_L - \frac{\mathbf{Q}_{21}^H \mathbf{Q}_{21}}{K^2}\right)^{-1}$.

The MSE of the LS estimator for the CIR from MS $1$ to antenna $a$
of BS $b$ can be derived as
\begin{align}\label{E:Apdx_MSE1_LS}
\mathrm{MSE}_{1,b,a}^{\mathrm{LS}}= \frac{\sigma_n^2}{p^uK}tr\{\mathbf{M}\}
=\frac{\sigma_n^2}{p^uK}tr\left\{\left(\mathbf{I}_L - \mathbf{U} \mathbf{\Lambda} \mathbf{U}^H\right)^{-1}\right\}
=\frac{\sigma_n^2}{p^uK} \sum_{l=0}^{L-1}{\frac{1}{1-\lambda_l^2}}.
\end{align}

The MSE of LS estimator for CIR from the MS $2$ can be derived in
the same way as
\begin{align}\label{E:Apdx_MSE2_LS}
\mathrm{MSE}_{2,b,a}^{\mathrm{LS}} &= \frac{\sigma_n^2}{p^uK}tr\left\{\frac{\mathbf{Q}_{21}}{K}\mathbf{M} \frac{\mathbf{Q}_{21}^H}{p^uK} + \mathbf{I}_L\right\} \nonumber\\
%&= \frac{\sigma_n^2}{p^uK}tr\{\frac{\mathbf{Q}_{21}}{K}(\mathbf{I}_L - \frac{\mathbf{Q}_{21}^H \mathbf{Q}_{21}}{K^2})^{-1} \frac{\mathbf{Q}_{21}^H}{K} + \mathbf{I}_L\} \nonumber\\
&= \frac{\sigma_n^2}{p^uK}tr\left\{\left(\mathbf{I}_L -
\frac{\mathbf{Q}_{21}^H \mathbf{Q}_{21}}{K^2}\right)^{-1}
\frac{\mathbf{Q}_{21}^H}{K}\frac{\mathbf{Q}_{21}}{K}\right\} +
\frac{\sigma_n^2}{K}tr\{\mathbf{I}_L\}
%& = \frac{\sigma_n^2}{p^uK}tr\{\mathbf{U}(\mathbf{I}_L -  \mathbf{\Lambda} )^{-1}\mathbf{U}^H \mathbf{U}\mathbf{\Lambda} \mathbf{U}^H\} + \frac{\sigma_n^2}{K}L
= \frac{\sigma_n^2}{p^uK} \sum_{l=0}^{L-1}{\frac{1}{1-\lambda_l^2}}.
\end{align}

Comparing (\ref{E:Apdx_MSE1_LS}) and (\ref{E:Apdx_MSE2_LS}), we
find that the MSEs of the LS estimator for the CIRs from different MSs are
identical.

\vspace{-0.4cm}
\section{Rate loss after taken expectation over
channel estimation errors} \label{A:proof-4} The average
interference power experienced by MS $m$ in (\ref{E:RateGap}) can be
derived by taking expectation with respect to the estimation
errors, i.e.,
\begin{align}\label{E:I_C}
\mathbb{E}\{I_m\} &= \mathbb{E}_{\mathbf{\tilde{g}}_m^f}
\left\{\sum_{j=1,j \neq m}^M  \left|(\mathbf{\tilde{g}}_m^f)^H \mathbf{v}_j\right|^2
\right\}
= \sum_{j=1,j \neq m}^M  \mathbb{E}_{\mathbf{\tilde{g}}_m^f}  \left\{ \left|\sum\nolimits_{b=1}^B (\mathbf{\tilde{g}}_{m,b}^f)^H \mathbf{v}_{b,j}\right|^2\right\} \nonumber \\
& = \sum_{j=1,j \neq m}^M \mathbb{E}_{\mathbf{\tilde{g}}_m^f}  \left\{\sum\nolimits_{b=1}^B\left| (\mathbf{\tilde{g}}_{m,b}^f)^H \mathbf{v}_{b,j}\right|^2 + 2 \Re \left\{\sum\nolimits_{t=1}^B\sum\nolimits_{u=1,u \neq t}^B (\mathbf{\tilde{g}}_{m,t}^f)^H \mathbf{v}_{t,j} (\mathbf{v}_{u,j})^H \mathbf{\tilde{g}}_{j,u}^f\right\} \right\}\nonumber\\
& \stackrel{(a)}= \sum_{j=1,j \neq m}^M  \sum_{b=1}^B
\mathbb{E}_{\mathbf{\tilde{g}}_m^f}
\left\{\left|(\mathbf{\tilde{g}}^f_{m,b})^H
\mathbf{v}_{b,j}\right|^2\right\}\nonumber\\
&\stackrel{(b)}= \sum_{j=1,j \neq m}^M \sum_{b=1}^B
\sigma_{e_{m,b}}^2 \|\mathbf{v}_{b,j}\|^2,
\end{align}
where $(a)$ follows because the channel estimation errors of the
channels from multiple BSs to one MS are assumed to be uncorrelated
and their expectations are zero. $(b)$ is derived by considering
that the covariance of estimation errors for the channels from
multiple antennas of one BS are the same, i.e., $\mathbb{E}
\{\mathbf{\tilde{g}}^{f}_{m,b} (\mathbf{\tilde{g}}^{f}_{m,b})^H\} =
\sigma^2_{e_{m,b}} \mathbf{I}_{N_t}$.

Substituting (\ref{E:I_C}) into (\ref{E:RateGap}), we can get the
upper bound of the rate loss under CoMP transmission as
\begin{align}
\Delta R_m^{\mathrm{UB}} &= \mathrm{log}_2\left[1 +\frac{
p^d}{\sigma_z^2}
\sum_{j=1,j \neq m}^M  \sum_{b=1}^B \sigma^2_{e_{m,b}} \|\mathbf{v}_{b,j}\|^2\right] \nonumber \\
&= \mathrm{log}_2\left[1 +
\sum_{j=1,j \neq m}^M  \sum_{b=1}^B \frac{\alpha^2_{m,b}
p^d}{\sigma_z^2}\frac{\sigma^2_{e_{m,b}}}
{\alpha^2_{m,b}} \|\mathbf{v}_{b,j}\|^2\right].
\end{align}

\bibliography{IEEEabrv,Houbib}

\begin{thebibliography}{10}
\providecommand{\url}[1]{#1}
\csname url@rmstyle\endcsname
\providecommand{\newblock}{\relax}
\providecommand{\bibinfo}[2]{#2}
\providecommand\BIBentrySTDinterwordspacing{\spaceskip=0pt\relax}
\providecommand\BIBentryALTinterwordstretchfactor{4}
\providecommand\BIBentryALTinterwordspacing{\spaceskip=\fontdimen2\font plus
\BIBentryALTinterwordstretchfactor\fontdimen3\font minus
  \fontdimen4\font\relax}
\providecommand\BIBforeignlanguage[2]{{%
\expandafter\ifx\csname l@#1\endcsname\relax
\typeout{** WARNING: IEEEtran.bst: No hyphenation pattern has been}%
\typeout{** loaded for the language `#1'. Using the pattern for}%
\typeout{** the default language instead.}%
\else
\language=\csname l@#1\endcsname
\fi
#2}}

\bibitem{Foschini-NetworkMIMO-06}
M.~K. Karakayali, G.~J. Foschini, and R.~A. Valenzuela, ``Network coordination
  for spectrally efficient communications in cellular systems,'' \emph{{IEEE}
  Wireless Commun. Mag.}, vol.~13, no.~4, pp. 56--61, Aug. 2006.

\bibitem{Tolli08}
{A. T\"olli, M. Codreanu, and M. Juntti}, ``Cooperative {MIMO-OFDM} cellular
  system with soft handover between distributed base station antennas,''
  \emph{{IEEE} Trans. Wireless Commun.}, vol.~7, no.~4, pp. 1428--1440, Apr.
  2008.

\bibitem{J.Jose-PilotContam-ISIT09}
J.~Jose, A.~Ashikhmin, T.~Marzetta, and S.~Vishwanath, ``Pilot contamination
  problem in multi-cell {TDD} systems,'' in \emph{Proc. IEEE Int. Symp.
  Information Theory (ISIT)}, Seoul, Korea, June 2009, pp. 2184--2188.

\bibitem{Kang-CellularPilot-Letter07}
G.~Kang, P.~Hasselbach, Y.~Yang, P.~Zhang, and A.~Klein, ``Pilot design for
  inter-cell interference mitigation in {MIMO OFDM} system,'' \emph{{IEEE}
  Commun. Lett.}, vol.~11, no.~3, pp. 237--239, Mar. 2007.

\bibitem{D.Katselis-TSunderCorrCH-08TSP}
D.~Katselis, E.~Kofidis, and S.~Theodoridis, ``On training optimization for
  estimation of correlated {MIMO} channels in the presence of multiuser
  interference,'' \emph{{IEEE} Trans. Signal Processing}, vol.~56, no.~10, pp.
  4892--4904, Oct. 2008.

\bibitem{S.Lee-VarPilotDens-TWC08}
S.~Lee, K.~Kwak, J.~Kim, and D.~Hong, ``Channel estimation approach with
  variable pilot density to mitigate interference over time-selective cellular
  {OFDM} systems,'' \emph{{IEEE} Trans. Wireless Commun.}, vol.~7, no.~7, pp.
  2694--2704, July 2008.

\bibitem{M.R.Raghavendra-InfRej-TVT09}
M.~R. Raghavendra, S.~Bhashyam, and K.~Giridhar, ``Interference rejection for
  parametric channel estimation in reuse-1 cellular {OFDM} systems,''
  \emph{{IEEE} Trans. Veh. Technol.}, vol.~58, no.~8, pp. 4342--4352, Oct.
  2009.

\bibitem{LiYe-OptimalTS-TC02}
Y.~Li, ``Simplified channel estimation for {OFDM} systems with multiple
  trasnmit antennas,'' \emph{{IEEE} Trans. Commun.}, vol.~1, no.~1, pp. 67--75,
  Jan. 2002.

\bibitem{CSISounding-ICC10}
H.~S. Kim, S.~H. Lee, and Y.~H. Lee, ``Channel sounding for multi-sector
  cooperative beamforming in {TDD-OFDM} wireless systems,'' in \emph{Proc. IEEE
  Int. Conf. Commun. (ICC)}, Cape Town, South Africa, May 2010.

\bibitem{A.Papadogiannis-DynamiCluster-08ICC}
A.~Papadogiannis, D.~Gesbert, and E.~Hardouin, ``A dynamic clustering approach
  in wireless networks with multi-cell cooperative processing,'' in \emph{Proc.
  IEEE Int. Conf. Commun. (ICC)}, Beijing, China, May 2008, pp. 4033--4037.

\bibitem{L.Thiele-VirtualPilots-VTC08}
M.~S. L.~Thiele, S.~Schiffermuller, V.~Jungnickel, and W.~Zirwas, ``Multi-cell
  channel estimation using virtual pilots,'' in \emph{Proc. IEEE Veh. Technol.
  Conf. Spring (VTC-S)}, Singapore, May 2008, pp. 1211--1215.

\bibitem{T.Kwon-PilotDelay-07VTC}
T.~Kwon, H.~Song, and D.~Hong, ``Robust channel estimation in multicell
  {OFDM(A)} downlink systems with propagation delay,'' in \emph{Proc. IEEE Veh.
  Technol. Conf. Spring (VTC-S)}, Dublin, Ireland, May 2007, pp. 1450--1454.

\bibitem{Zhang-Molish-async08}
H.~Zhang, N.~B. Mehta, A.~F. Molisch, J.~Zhang, and H.~Dai, ``Asynchronous
  interference mitigation in cooperative base station systems,'' \emph{{IEEE}
  Trans. Wireless Commun.}, vol.~7, no.~1, pp. 155--165, Jan. 2008.

\bibitem{HY09}
S.~Han, C.~Yang, M.~Bengtsson, and A.~I. Perez-Neira, ``Channel norm based user
  scheduling in coordinated multi-point systems,'' in \emph{Proc. IEEE Glob.
  Telecom. Conf. (GlobeCom)}, Nov. 2009.

\bibitem{RG05}
M.~R. Raghavendra and K.~Giridhar, ``Improving channel estimation in {OFDM}
  systems for sparse multipath channels,'' \emph{{IEEE} Signal Processing
  Lett.}, vol.~12, no.~1, pp. 52--55, Jan. 2005.

\bibitem{MYi-EstPDP-09TVT}
K.~C. Hung and D.~W. Lin, ``Pilot-based {LMMSE} channel estimation for {OFDM}
  systems with power-delay profile approximation,'' \emph{{IEEE} Trans. Veh.
  Technol.}, vol.~59, no.~1, pp. 150--159, Jan. 2009.

\bibitem{LiYe-RobustCE-TC98}
Y.~Li, L.~J. Cimini, and N.~R. Sollenberger, ``Robust channel estimation for
  {OFDM} systems with rapid dispersive fading channels,'' \emph{{IEEE} Trans.
  Commun.}, vol.~46, no.~7, pp. 902--915, July 1998.

\bibitem{MatrCompu}
G.~H. Golub and C.~F.~V. Loan, \emph{Matrix Computations}.\hskip 1em plus 0.5em
  minus 0.4em\relax Baltimore MD: Johns Hopkins University Press, 1996.

\bibitem{I.Barhumi-OptimalPilot-TSP03}
I.~Barhumi, G.~Leus, and M.~Moonen, ``Optimal training design for {MIMO OFDM}
  systems in mobile wireless channels,'' \emph{{IEEE} Trans. Signal
  Processing}, vol.~51, no.~6, pp. 1615--1624, June 2003.

\bibitem{LTE-TR36.211}
{3GPP Long Term Evolution (LTE)}, ``Physical channels and modulation,''
  \emph{TSG RAN TR 36.211 v8.4.0}, Sept. 2008.

\bibitem{Huawei-SRS-58}
Y.~Ogawa, T.~Takata, T.~Iwai, D.~Imamura, K.~Hiramatsu, and K.~Miya, ``Pilot
  signal generation scheme using frequency dependent cyclic shift sequence for
  inter-cell interference mitigation,'' in \emph{Proc. IEEE Radio and Wireless
  Symposium (RWS)}, San Diego, CA, Jan. 2009, pp. 421--424.

\bibitem{Caire09}
G.~Caire, N.~Jindal, M.~Kobayashi, and N.~Ravindran, ``Multiuser {MIMO}
  achievable rates with downlink training and channel state feedback,''
  \emph{{IEEE} Trans. Inform. Theory}, vol.~56, no.~6, pp. 2845--2866, June
  2010.

\bibitem{ZCseq}
D.~C. Chu, ``Polyphase codes with good periodic correlation properties,''
  \emph{{IEEE} Trans. Inform. Theory}, vol.~18, no.~4, pp. 531--532, July 1972.

\end{thebibliography}

\newpage

\end{document}